\documentclass[twocolumn]{aa}
\usepackage{graphicx,natbib,textcomp}
\usepackage{txfonts}
\begin{document}
   \title{A dual emission mechanism in \object{Sgr A*/L$^\prime$} ?\thanks{Based on observations collected at the ESO VLT Yepun telescope, proposal 073.B-0665}}

   \author{Y. Cl\'enet\inst{1}
          \and
          D. Rouan\inst{1}
          \and
          D. Gratadour\inst{1}
          \and
          O. Marco\inst{2}
          \and
          P. L\'ena\inst{1}
          \and
          N. Ageorges\inst{2}
          \and
          E. Gendron\inst{1}
          }

   \offprints{Y. Cl\'enet}

   \institute{ Observatoire de Paris, LESIA, 5 place Jules Janssen, F-92195 Meudon Cedex, France\\
              \email{firstname.lastname@obspm.fr}
         \and
             European Southern Observatory (ESO), Alonso de Cordova 3107, Vitacura, Casilla 19001, Santiago 19, Chile\\
             \email{nageorges@eso.org, omarco@eso.org}
             }

   \date{Received ; accepted }

   \abstract{We have collected in 2004 adaptive optics corrected L$^\prime$ images of the Galactic Center region with NAOS-CONICA at VLT. A strong variability was observed as well as a correlation between the photocenter positions and fluxes of the L$^\prime$-band counterpart of \object{Sgr~A*}. It is interpreted as the combined emission of a point-like flaring source at the position of \object{Sgr~A*/IR} itself and an extended dust structure, 75 mas south west of \object{Sgr~A*/IR}, which we name \object{Sgr~A*-f}. We examine the different possible mechanisms to explain this dual \object{Sgr~A*} L$^\prime$ emission and conclude it is likely a flaring emission associated to energetic events in the close environment of the black hole plus a quiescent emission resulting from the collision of \object{Sgr~A*-f} by a jet from \object{Sgr~A*}.
      
   }

   \maketitle
%

\section{Introduction}
Since the advent of adaptive optics (AO) at VLT and Keck, the infrared (IR) counterpart of \object{Sgr~A*} is routinely observed from the H- (1.65 $\mu$m) to the M$^\prime$-band (4.8 $\mu$m) \citep{genzel03,ghez04,clenet04a,clenet04b} and has demonstrated a high variability at different time scales:

-- H-, Ks-, L$^\prime$-band flares, with typically a 1 h duration and a 1.5 to 5 flux magnification \citep{genzel03}, similar in time scales to the X-ray flares \citep{baganoff01,porquet03,eckart04};

-- L$^\prime$-, M$^\prime$-band variations on daily time scales \citep{ghez04,clenet04b}, with a flux magnification up to~5;

-- L$^\prime$-band variations on yearly time scales \citep{clenet04b}, with a flux magnification between 3 and 5.

Even though the observed IR fluxes are much below the predictions for a 3$\times$10$^6$~M$_\odot$ black hole radiating at the Eddington rate, they can be well fitted by a non thermal electron emission, whatever is the still unknown source accelerating these electrons:  a jet \citep{markoff01,yuan02} or the accretion flow \citep{yuan03,yuan04}.

We report the detection of the L$^\prime$-band counterpart of \object{Sgr~A*} (hereafter \object{Sgr~A*/L$^\prime$}) as a source whose intensity and position are time dependent. It results from the combined emission of  \object{Sgr~A*/IR} itself and a nearby extended structure we name \object{Sgr~A*-f}. We first detail the observations and data reduction, next explain the photometry and astrometry methods, then examine the position-flux correlation of \object{Sgr~A*/L$^\prime$} and finally explore different mechanisms for this \object{Sgr~A*/L$^\prime$} emission.

\section{Observations and data reduction}
We have performed L$^\prime$-band (3.8 $\mu$m, 0.0271\arcsec/pixel) observations of the Galactic Center region during five 2004 nights (Apr 24, Apr 25, June 13, Aug 10 and Sept 20, hereafter Night 1, 2, 3, 4 and 5 respectively) with NACO \citep{lenzen98,rousset00}, the AO assisted imager installed at the 8m VLT UT4 telescope. 

On-source (A) and on-sky (B) 1024$\times$1024 frames, each one resulting from the mean of 60 subintegrations of 0.175 s, have been alternatively acquired following an ABBA cycle, with a random jitter within a box   6\arcsec\ wide every other cycle. The sky position was 3\arcmin\ away from the on-source position. The total on-source integration times were 27.3 min, 16.8 min, 49 min, 10.5 min and 9.8 min for Nights~1,~2,~3,~4,~5 respectively.

After sky subtraction, flat fielding and bad pixel correction, the recentering  has been done in two steps: (1) the relative offsets between individual images have been estimated by computing their cross-correlation function, leading to an accuracy of $\approx$0.5 pixel; (2) a subpixel recentering has been performed with a maximum likelihood based algorithm \citep{gratadour05}, leading to a final accuracy of $\approx$0.01 pixel i.e. $\approx$0.3 mas.

For each night, to allow for a study of the time evolution of the \object{Sgr~A*/L$^\prime$} flux with a sufficient signal to noise, we have averaged the recentered individual images into groups of four, leading to a cube of averaged images with a typical sampling of $\approx$265 s and a 42 s integration time for each averaged image.

\section{Astrometry and photometry}
\begin{figure}
\centering
\resizebox{\hsize}{!}{\includegraphics{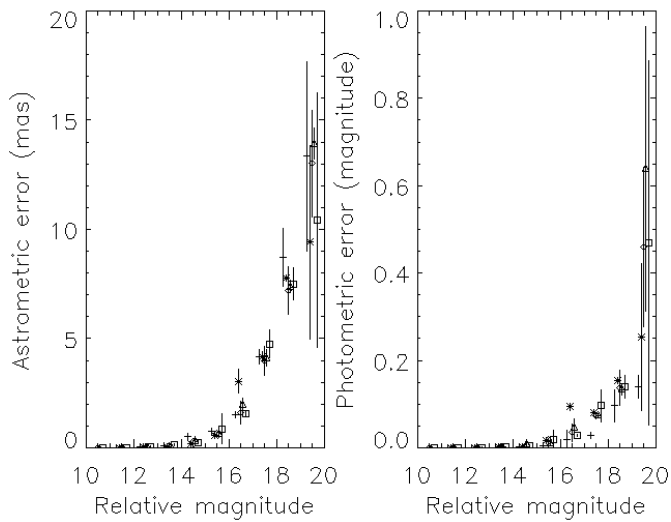}}
\vspace{-4mm}
\caption{Astrometric/photometric errors and their standard deviations. Crosses are for Night1, stars for Night2, diamonds for Night3, triangles for Night4, and squares for Night5. Errors are estimated for each magnitude bin of the luminosity function computed by StarFinder. The translation to get the absolute photometry can be done with the zero-point of each night ($\approx$4.5). To avoid confusion, the different symbols are horizontally shifted with respect to the bin centers.}
\label{fig:error}
\vspace{-3mm}
\end{figure}

The astrometry and the photometry of all averaged images have been obtained with StarFinder  \citep{diolaiti00}, a PSF-fitting software designed for AO data. 

To estimate the  StarFinder accuracy for a given averaged image,  we have added, in each magnitude bin of the luminosity function computed by StarFinder, 20\% of synthetic stars at random positions and with random magnitudes and re-analyzed this new image with StarFinder. For each night, this procedure has been applied to three different averaged images. 

Up to the relative magnitude bin [18,19] (L$^\prime$$\approx$[13.5,14.5]), the error dispersion between the different nights is small, as well as the standard deviation associated to each night error value (Fig.~\ref{fig:error}). This is no longer the case for the last bin [19,20] (L$^\prime$$\approx$[14.5,15.5]), most probably because of the lower accuracy of StarFinder for these fainter stars.

The absolute photometry has been computed as in \citet{clenet04b}, using the photometry of the non variable stars \object{IRS~16C}, \object{IRS~29N} and \object{IRS~33SE} published in \citet{blum96}. Given the accuracy of this zero-point computation ($\approx$0.08 mag) and the StarFinder accuracy (Fig.~\ref{fig:error}), we finally estimate the photometric error for the L$^\prime$$\approx$14 stars to be 0.18 mag for Night1, 0.17 mag for Night2, and 0.16 mag for Night3, Night4 and Night5. 

Dereddened fluxes have been computed as in \citet{clenet04b} with an L$^\prime$-band extinction A$_\mathrm{L^\prime}$=1.30 and a zero magnitude value F$_0$(L$^\prime$)=248~Jy.

Following \citet{clenet04b}, the astrometry has been calculated with respect to \object{Sgr~A*} taken as the dynamical center (DC) of the \object{S2} star orbit. The offsets between \object{S2} and DC are given in Table~\ref{table:offset} and the final astrometric errors in Table~\ref{table:astromerror}. 

\begin{table}
\caption{Offsets between \object{S2} and the dynamical center \object{Sgr~A*}, computed with the \object{S2} orbit parameters of \citet{eisenhauer03}. Errors are computed by simulating several \object{S2} orbits with gaussian distributions of the orbit parameters (cf.~\citealt{clenet04b}). }             
\label{table:offset}     
\centering                          
\begin{tabular}{c c c c c c}        
\hline\hline                 
Date & Apr. 24 & Apr. 25 & Jun. 13 & Aug. 10 & Sep. 20 \\    
\hline                        
$\Delta \alpha$ (mas) & 31$\pm$5 &  31$\pm$5 & 30$\pm$5 & 29$\pm$5 & 28$\pm$5 \\      
$\Delta \delta$ (mas) & 111$\pm$9 & 111$\pm$9 & 115$\pm$8 & 120$\pm$5 & 123$\pm$4 \\
\hline                                   
\end{tabular}
\end{table}

\begin{table}
\caption{Astrometric errors for the L$^\prime$$\approx$14 stars, given the DC positioning error (Table~\ref{table:offset}) and the StarFinder accuracy (Fig.~\ref{fig:error}).}             
\label{table:astromerror}     
\centering                          
\begin{tabular}{c c c c c c}        
\hline\hline                 
Date & Apr. 24 & Apr. 25 & Jun. 13 & Aug. 10 & Sep. 20 \\    
\hline                        
$\sigma_\alpha$ (mas) & 8 &  8 & 8 & 8 & 8 \\      
$\sigma_\delta$ (mas) & 11 & 11 & 10 & 8 & 7 \\
\hline                                   
\end{tabular}
\end{table}

\begin{figure*}
\centering
\resizebox{\hsize}{!}{\includegraphics[angle=90]{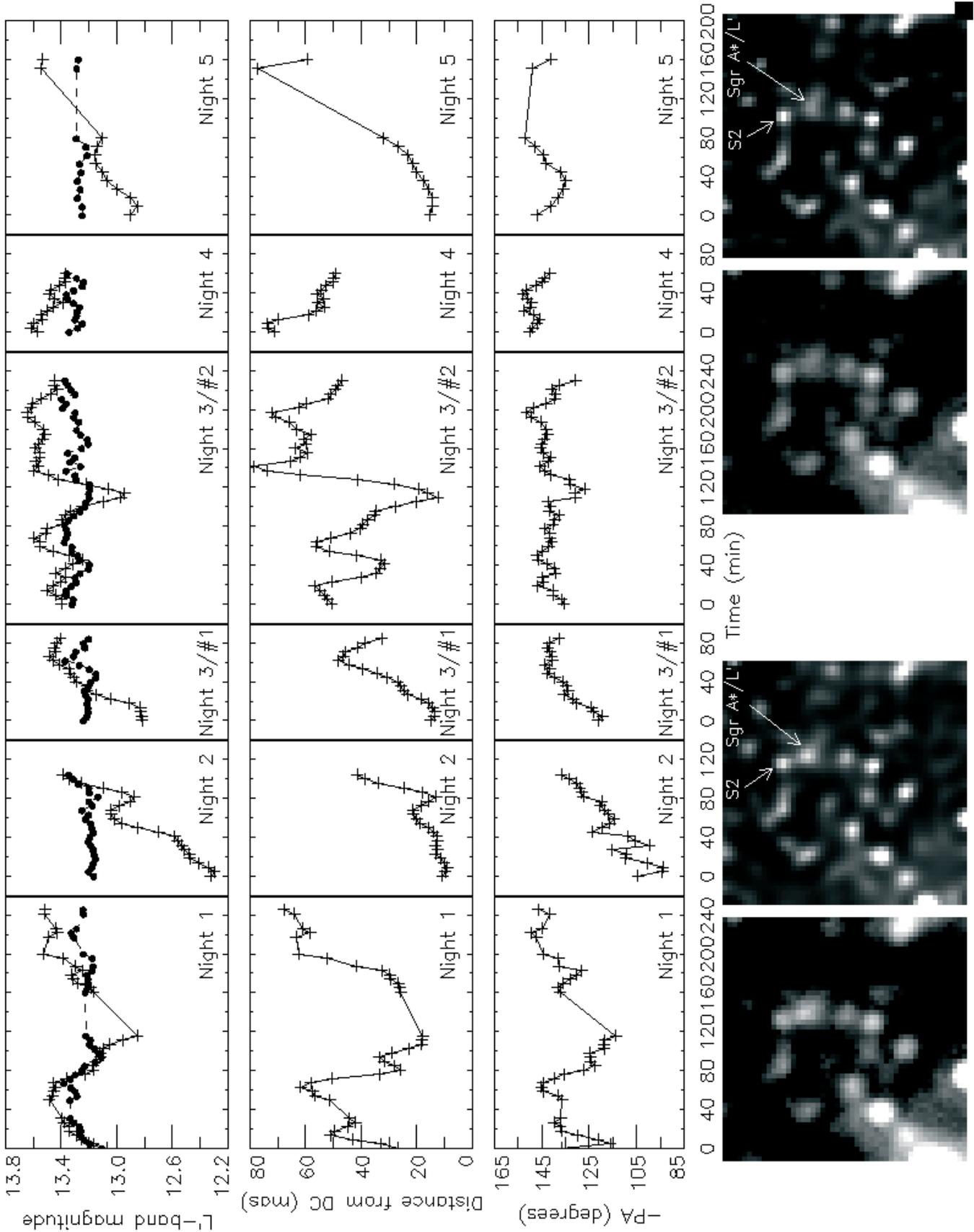}}
      \caption{Upper line: \object{Sgr A*/L$^\prime$} (continuous line/crosses) and \object{S2} (dash line/filled circles) light curves. Second line: distance in mas between  \object{Sgr~A*/L$^\prime$} and DC. Third line: Opposite of the \object{Sgr~A*/L$^\prime$} polar angle with respect to DC. The curves of the different nights are separated by a vertical line. All curves have been smoothed with a boxcar average of 2. Bottom line:1.4\arcsec$\times$1.4\arcsec\ Night 3 images of the central stellar cluster during \object{Sgr~A*/L$^\prime$} flare (couple of images on the left) and quiescent (couple of images on the right) states. Each couple of images is made of a 105 s integration time reduced image and the deconvolved corresponding one, obtained with 40 iterations of the maximum likelihood deconvolution IDL procedure. The deconvolved images are presented here only to highlight the sources position/extension and have not been used elsewhere for any quantitative analysis.}
         \label{fig:night}
   \end{figure*}

\begin{figure*}
\centering
\resizebox{\hsize}{!}{\includegraphics{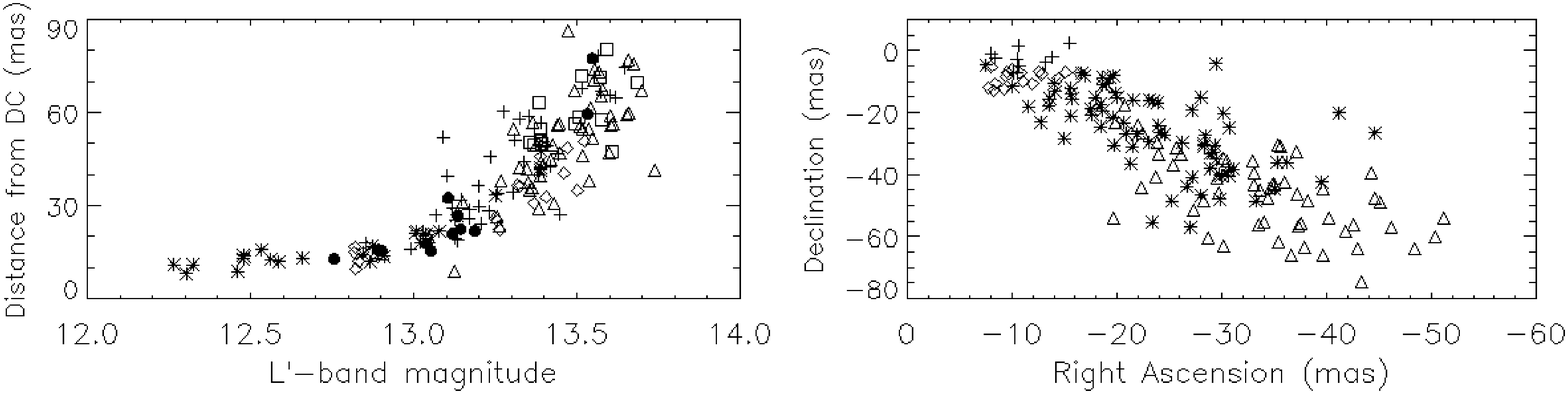}}
      \caption{Left: Distance of  \object{Sgr~A*/L$^\prime$} from DC in mas as a function of the \object{Sgr~A*/L$^\prime$} magnitude. Crosses, stars, diamonds, triangles, squares and filled circles are Night1, 2, 3/part1, 3/part2, 4, 5 data points respectively. Right: \object{Sgr~A*/L$^\prime$} positions with respect to DC. Crosses represent positions for  12.2$\leq$L$^\prime$(\object{Sgr~A*/L$^\prime$})$<$12.6, diamonds for 12.6$\leq$L$^\prime$(\object{Sgr~A*/L$^\prime$})$<$13.0, stars for 13.0$\leq$L$^\prime$(\object{Sgr~A*/L$^\prime$})$<$13.4, triangles for 13.4$\leq$L$^\prime$(\object{Sgr~A*/L$^\prime$})$<$13.8. }
\label{fig:correlation}
\end{figure*}

\section{Results}

The photocenter positions and fluxes of \object{Sgr~A*/L$^\prime$}  were measured on each averaged image using StarFinder. Firstly, as described in \citet{clenet04b}, \object{Sgr~A*/L$^\prime$} has exhibited variability on time scales of about one hour: flares with an amplification factor $\approx$2 during Night 1 and Night 3/part2; only decays or rises, with an amplification factor $\approx$3, for the other nights.  

Secondly, we confirm the offset of \object{Sgr~A*/L$^\prime$} in quiescence with respect to DC reported in \citet{clenet04b} at L$^\prime$ and M$^\prime$; we also observe a clear correlation between the \object{Sgr~A*/L$^\prime$} flux and a) the DC-\object{Sgr~A*/L$^\prime$} distance, b) the \object{Sgr~A*/L$^\prime$} position angle (PA) with respect to DC. We show that these correlations are reproducible from one night to another: \object{Sgr~A*/L$^\prime$} and DC are all the more distant and \object{Sgr~A*/L$^\prime$} is all the more shifting toward DC than \object{Sgr~A*/L$^\prime$} is bright (Fig.~\ref{fig:night}, ~\ref{fig:correlation}). 

Hence, our observations point at a dual mechanism responsible for the 3-5 $\mu$m emission in \object{Sgr~A*}:  one traced by the flaring state, unresolved and associated to \object{Sgr~A*/IR} itself; another one associated to the quiet state, clearly resolved with a typical extension of about 130 mas, occurring at the south-west of the dynamical centre (PA$\approx$-130\degr) and at a distance of 75 mas typically (600 AU) from it. A simple argument on the light velocity limit comforts the likeliness of this dual mechanism from two distinct sources where the emission originates from two distinct sources that NACO has marginally separated spatially, thanks to the extension of its quiescent part.

\section{Discussion}

The flaring source compactness and the typical variation times measured at L$^\prime$ do not conflict with the interpretation of \citet{eckart04}: both near-IR and X- emissions would be related to synchrotron self-Compton emission of electrons. The flare could be mainly due to Doppler boosted emission from material moving toward the observer, either matter at the base of a jet, or matter bound to the black hole on one of the innermost stable orbits. No simultaneous observations at several IR wavelengths have yet been done  to allow one to derive the IR spectral index and  we can only guess that flares are rather reproducible phenomena, as indicated by the similar behaviour of rises and decays in the observed L$^\prime$-band events.

Though, another process should be invoked to explain a quiescent emission as far as 600 AU, corresponding to about 10$^4$ Schwarzschild radius ($R_\mathrm{S}$).  A stellar emission could hardly account for this quiescent emission since: a) no K-band source can be observed at this location in the images of  \citet{eckart04} (see their Figure~2), b) the inferred K-L$^\prime$ color (larger than 3) would be too large, c) the source appears extended.

An estimate of the IR color temperature $T_\mathrm{c}$ corresponding to the  \object{Sgr~A*/L$^\prime$} quiescent emission can be computed by assuming the quiescent K-band dereddened flux density found by \citet{eckart04} ($F_\nu$(K)=1.9 mJy), a quiescent L$^\prime$-band dereddened flux density of 3.0 mJy and by resolving the optically thin case equation:

\begin{center}
$F_\nu(K)/F_\nu(L^\prime)=Q_\mathrm{abs}(K)/Q_\mathrm{abs}(L^\prime)\times B_\nu(T_\mathrm{c},K)/B_\nu(T_\mathrm{c},L^\prime)$
\end{center}

where $Q_\mathrm{abs}$ is the absorption cross section coefficient and $B_\nu(T_\mathrm{c},\lambda)$ the Planck function at the temperature $T_\mathrm{c}$ and the wavelength $\lambda$. From the dust grain properties of \citet{laor93} for dust grain sizes ranging from 0.01 to 0.1 $\mu$m, we obtain a color temperature between 1110 K and 1130 K for silicate grains, between 870 K and 930 K for graphite grains.

To estimate a maximum equilibrium temperature $T_\mathrm{e}$ of dust grains heated by \object{Sgr~A*} itself,  we equal the powers received and radiated by the grain:

\begin{center}
$\displaystyle{\int_{0}^{\infty}\frac{L_\lambda}{4\pi r^2}}\,\pi a^2\, Q_\mathrm{abs}(\lambda)\,d\lambda=\displaystyle{\int_{0}^{\infty}}4\pi a^2\,Q_\mathrm{abs}(\lambda)\,\pi B_\lambda(T_e,\lambda)\,d\lambda$
\end{center}
 
where $L_\lambda$ is the \object{Sgr~A*} flux density, $a$ the dust grain size and $r$ the distance we have found between the central black hole and  \object{Sgr~A*/L$^\prime$} in its quiescent state ($r$=600 UA, ie about 75 mas). Since the smallest wavelengths dominate the first term, we approximate $L_\lambda$ by the \object{Sgr~A*} quiescent state power law found by \citet{baganoff01} and usually adopted to constrain the \object{Sgr~A*} emission models:

\begin{center}
$L_\lambda=7.1\times 10^{-35}\left(\displaystyle{\frac{\lambda}{3.1\times 10^{-10}}}\right)^{-0.8}$
\end{center}

For dust grain sizes ranging from 0.001 to 0.1 $\mu$m, we obtain an equilibrium temperature between 70 K and 90 K for silicate grains, between 80 K and 110 K for graphite grains. About one order of magnitude is then found between the quiescent IR color temperature $T_\mathrm{c}$ of \object{Sgr~A*/L$^\prime$} and the temperature $T_\mathrm{e}$ expected if the accretion disk emission was the dust heating source at the \object{Sgr~A*/L$^\prime$} location. This latter process is therefore excluded  to explain the L$^\prime$ quiescent emission.

We have also explored the possibility that the quiescent extended L$^\prime$ emission could be due to synchrotron emission from a jet. The absorption coefficient of the synchrotron emission is $ \kappa_{\nu} = 0.019\, (3.5\times 10^9)^p N_0 B^{(p+2)/2} \nu^{-(p+4)/2} $, where $p$ and $N_0$ describe the power-law dependence of the electrons energy distribution ($n_e(E) = N_0 E^{-p}$) and $B$ is the magnetic field. Using the jet model described in \citet{falcke00}, completed by the analytical expression of the quantity $\gamma_{j}\, \beta_{j}$ established in \citet{falcke96}, we have set the parameters at the jet nozzle according to  \citet{yuan02}, i.e. so that to fit the  millimetric observations. As expected, the resulting  optical depth at a typical distance of 600 AU from the black hole is extremely low, so that the quiescent extended L$^\prime$ emission cannot be explained by the high frequency wing of synchrotron radiation from a jet. 

If we now consider the mass loss rate in the jet found by \citet{yuan02}, $\dot{M}_{jet}$=4.3$\times$10$^{-9}$ M$_\odot$yr$^{-1}$, and translate it in kinetic power delivered by the jet (assuming an equipartition between ions and electrons), we get $P_{jet}$=$\dot{M}_{jet}\times m_e/m_p\times (\beta c)^2$, where $m_e$, $m_p$, $\beta$ are the mass of electron, the mass of  proton and the electron to light speed ratio, respectively. With $\beta$=0.95 \citep{yuan02}, we obtain $P_{jet}$=1.2$\times$10$^{35}$ erg/s=32 L$_\odot$.

On the other hand, the total luminosity, if due to grains, that would correspond to the observed L$^\prime$-band luminosity $L_{L^\prime}$ is: 

\begin{center}
$L_{tot}=\frac{\displaystyle{\int_{0}^{\infty}Q_\mathrm{abs}(\lambda) B_\lambda(T_c,\lambda)\,d\lambda}}{\displaystyle{\int_{L^\prime}Q_\mathrm{abs}(\lambda) B_\lambda(T_c,\lambda)\,d\lambda}}\times L_{L^\prime}$
\end{center}

For silicate grains at $T_c$=1120 K and graphite grains at $T_c$=900 K, whatever the grain size from from 0.001 to 0.1 $\mu$m, it comes $L_{tot}$=12 L$_\odot$ and $L_{tot}$=6 L$_\odot$ respectively, which are of the same magnitude order as the kinetic power delivered by the jet $P_{jet}$. Therefore, one could consider the energy transfer of this jet power to the dusty material surrounding \object{Sgr A*} through collisions as a possible mechanism to explain the \object{Sgr~A*/L$^\prime$} quiescent emission. 

In this description, the observed quiescent state extended source would  be a small dust condensation, close to \object{Sgr~A*} and heated through a colliding jet. The observed position-magnitude correlation would then result from Starfinder measuring the position/magnitude of the combined emission of \object{Sgr~A*/IR} itself and this dust structure. 

We note that the \object{Sgr~A*/IR} naming should be reserved to the flaring location since the role of the black hole  is obvious for this state. We propose to name \object{Sgr~A*-f} (standing for \object{Sgr~ A*} "flake") the extended emission located south west of \object{Sgr~A*/IR}. \object{Sgr~A*/L$^\prime$} is then the observed combined emission of \object{Sgr~A*/IR} and \object{Sgr~A*-f}.

We cannot totally exclude that \object{Sgr~A*-f} belongs to the mini-spiral and is then located at a fraction of parsec from \object{Sgr~A*} along the line of sight. Though, the physical association of \object{Sgr~A*/IR} and \object{Sgr~A*-f} is supported by the presence of a chain of gaseous clouds, probably associated to \object{Sgr~A*} through a ridge of emission, observed in the radio continuum between \object{Sgr~A} and the mini-cavity \citep{yusef90}. These clouds would be driven by fast winds towards the gas streamers orbiting  \object{Sgr~A*} (the "Minispiral") and create by shocks the mini-cavity. 

The origin of these radio features is still discussed: \citet{wardle92} and \citet{melia96} proposed they are formed by the gravitational collimation of the \object{IRS~16} cluster stellar winds by \object{Sgr~A*}. Alternatively, \citet{lutz93} invoked a jet from \object{Sgr~A*}, similarly to our scenario, as the wind source.

\begin{figure}
\centering
\resizebox{\hsize}{!}{\includegraphics{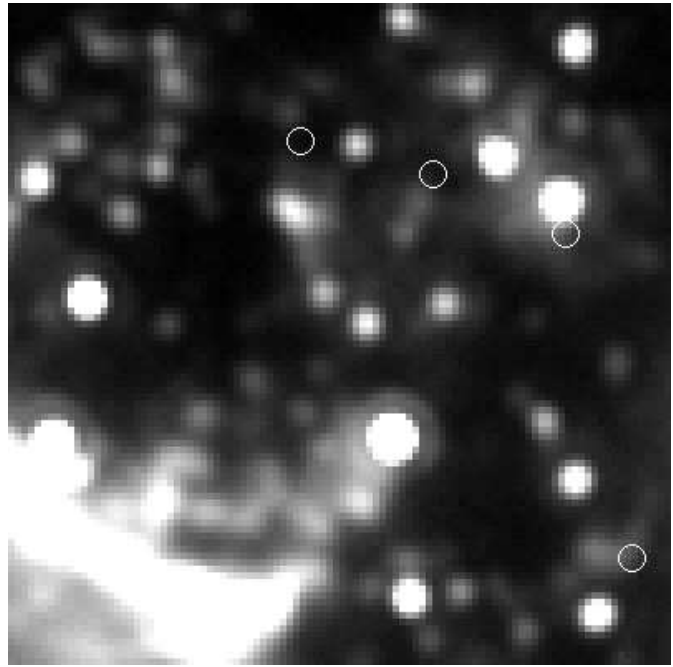}}
\caption{Encircled \ion{H}{ii} components positions on a 2.4\arcsec$\times$2.4\arcsec\ L$^\prime$ NACO image. From left to right: \ion{H}{ii} 5D, \ion{H}{ii} 5C, source $\epsilon$, source $\xi$ \citep{yusef98,zhao99}. The circle diameter is the 0.1\arcsec\ radio resolution. ID18 is the bright extended source just south to the \ion{H}{ii} 5D circle.}
\label{fig:blob}
\end{figure}

The position of these gaseous \ion{H}{ii} components, taking into account their estimated proper motions  and their fairly extended sizes \citep{yusef98,zhao99}, is very close to very red IR extended sources (Fig.~\ref{fig:blob}), e.g. \ion{H}{ii} 5D in \citet{zhao99} and ID18 in \citet{clenet04a}. This suggests a common origin, if not an identification, between the sources at the different wavelengths. \object{Sgr~A*-f} could then related to this chain of structures, possibly to the unnamed feature in Fig.~3 upper panel of \citet{zhao99}, located, in 1998, 0.35\arcsec\ south west to \object{Sgr~A*}: a proper motion similar to the other \ion{H}{ii} structures ($\approx$25 mas yr$^{-1}$) could have shifted this feature very close to the 2004 position of \object{Sgr~A*-f}.

In the future, to constrain the aforementioned physical model of a jet interacting with Sgr A*-f, simultaneous observations at different IR wavelengths of \object{Sgr~A*/IR} in both quiescent and flare states will be essential. An instrument such as the NACO prism disperser, which covers simultaneously the IR spectrum from J (1.3 $\mu$m) to M (4.8 $\mu$m), could be efficiently dedicated to this task. 

\end{document}